\documentclass[twocolumn]{revtex4-1}          
\pdfoutput=1
\usepackage{graphicx, color}
\usepackage{amsmath, amssymb, amsfonts, mathrsfs}
\usepackage{times}
\usepackage{latexsym, color, amssymb}
\newcommand{\ud}{\mathrm{d}}

\newcommand{\boltzmann}{k_{\textrm{B}}}
\newcommand{\functional}{$\mathscr{F}[\{\omega_{k}\}]$ }
\newcommand{\omegaS}{$\omega(\mathbf{s})$ }

\def\beq{\begin{equation}}
\def\eeq{\end{equation}}

\begin{document}


\title{Machine Learning for Parameter Auto-tuning in Molecular Dynamics Simulations: Efficient Dynamics of Ions near Polarizable Nanoparticles}

\author{JCS Kadupitiya}
\affiliation{Intelligent Systems Engineering, Indiana University, 700 N. Woodlawn Avenue, Bloomington, Indiana 47408, USA}
\author{Geoffrey C. Fox}
\affiliation{Intelligent Systems Engineering, Indiana University, 700 N. Woodlawn Avenue, Bloomington, Indiana 47408, USA}
\author{Vikram Jadhao}
\email{vjadhao@iu.edu}
\affiliation{Intelligent Systems Engineering, Indiana University, 700 N. Woodlawn Avenue, Bloomington, Indiana 47408, USA}


\begin{abstract}
Simulating the dynamics of ions near polarizable nanoparticles (NPs) using coarse-grained models is extremely challenging due to the need to solve the Poisson equation at every simulation timestep. Recently, a molecular dynamics (MD) method based on a dynamical optimization framework bypassed this obstacle by representing the polarization charge density as virtual dynamic variables, and evolving them in parallel with the physical dynamics of ions. We highlight the computational gains accessible with the integration of machine learning (ML) methods for parameter prediction in MD simulations by demonstrating how they were realized in MD simulations of ions near polarizable NPs. An artificial neural network based regression model was integrated with MD simulation and predicted the optimal simulation timestep and optimization parameters characterizing the virtual system with $94.3\%$ success. 
The ML-enabled auto-tuning of parameters generated accurate dynamics of ions for $\approx 10$ million steps while improving the stability of the simulation by over an order of magnitude. The integration of ML-enhanced framework with hybrid OpenMP/MPI parallelization techniques reduced the computational time of simulating systems with thousands of ions and induced charges from thousands of hours to tens of hours, yielding a maximum speedup of $\approx 3$ from ML-only acceleration and a maximum speedup of $\approx 600$ from the combination of ML and parallel computing methods. Extraction of ionic structure in concentrated electrolytes near oil-water emulsions demonstrates the success of the method. The approach can be generalized to select optimal parameters in other MD applications and energy minimization problems.
\end{abstract}


\maketitle

\section{Introduction}
Many biological and synthetic nanoparticle (NP) systems are polarized in the presence of electric fields generated by surrounding ions and other macromolecular charged species \citep{levin1, clapham, henderson}. Examples include proteins and DNA in an aqueous cellular medium, emulsions where oil and water are partitioned, and gold NPs dispersed in water. Accurate knowledge of ionic structure near the surface of these NPs enables the understanding of many nanoscale phenomena associated with these materials such as protein conformational changes \citep{honig}, DNA precipitation \citep{raspaud}, spontaneous emulsification \citep{sacanna}, and NP self-assembly \citep{levin1}. Extracting this structure by simulating the dynamics of ions in the presence of polarizable NPs using coarse-grained models is challenging due to the need to compute polarization (induced) charges in order to propagate the ion configuration \citep{allen1,marchi,santos,holm2014}. This computation typically involves solving the  second-order Poisson differential equation in 3-dimensional space at each simulation timestep, making the use of conventional nanoscale simulation methods very time consuming and inefficient. Because of these computational challenges, the problem of extracting ionic structure near polarizable NPs has been a subject of intense research \citep{marchi,boda,allen1,santos,jso1,jso2,holm2014,luijten2015,qin2016image,netz2018}. 

The problem is often re-casted in terms of energy minimization for which different candidate functionals and associated minimization methods have been proposed \citep{marchi,allen1,attard,luijten.jcp2014}. Among these techniques, a molecular dynamics (MD) method based on the dynamical optimization of an energy functional enabled the replacement of the expensive solution of the Poisson equation at each simulation step with an on-the-fly computation of surface polarization charges \citep{jso1,jso2,jing2015}. The main focus of this paper is to highlight the computational gains accessible by integrating machine learning (ML) methods for parameter auto-tuning in MD simulations by demonstrating how these gains were realized in the MD simulations of ions near polarizable NPs based on the dynamical optimization framework.

In the dynamical optimization framework, inspired by the Car-Parrinello method for simulating ion-electron systems \citep{car-parrinello}, an energy functional of the induced charge density is dynamically optimized resulting in the physical dynamics of ions in parallel with the update of the virtual variables characterizing the induced charge density. The virtual system is evolved in a manner that keeps the induced charges close to the free-energy minimum (``ground state'') corresponding to the evolving ionic configuration. The advantages associated with the on-the-fly computation of polarization effects in conjunction with the reduction in computational costs achieved by solving for the scalar induced charge density variable have enabled the study of electrolyte solutions near polarizable NPs using this framework \citep{jso1,jing2015}. However, the applicability of the original method is limited by the absence of a framework that automates the process of selecting the ``good'' parameters characterizing the virtual system as well as the optimal simulation timestep. These quantities determine the stability, accuracy, and overall efficiency of the dynamical optimization framework and they are found by a tedious process of trial and error that is informed by domain experience. Further, these parameters are selected at the start of the simulation and held fixed throughout the simulation, to often relatively conservative values, in order to ensure the long-time stability of the dynamics of ions. 

Recent years have witnessed a remarkable growth in the use of ML to enhance computational methods aimed at understanding  phenomena in materials science, biology, neuroscience, and physics \citep{ml.atomic2017,melko2017,balakrishnan2005multilayer,sam2017,fu2017,long2015machine, ferguson2017machine,ward2018matminer}. ML has been applied to identify interesting parameter spaces \citep{glotzer2017}, predict parameters \citep{balakrishnan2005multilayer}, update configurations \citep{botu2015adaptive}, infer assembly landscapes \citep{long2015machine,ferguson2017machine}, predict properties of materials \citep{ward2016general,ward2018machine} and classify phases of matter \citep{melko2017}.
Inspired by these recent developments, we describe an approach to integrate ML and MD methods to predict and auto-tune relevant parameters and simulation timestep. This approach is applied to the dynamical optimization framework to predict on-the-fly the virtual system parameters and simulation timestep that keep the polarization charge density close to the ground state determined by the evolving ionic configuration at all times during the simulation. The demonstration of the use of ML to predict and tune the MD simulation timestep has broad applicability. Similarly, we expect that the idea of using ML for predicting the virtual system parameters can be extended to enhance the original Car-Parrinello molecular dynamics techniques \citep{car-parrinello} for simulating ion-electron systems.

The use of ML to enhance the performance of the dynamical optimization framework is demonstrated using an $O(n^2)$ algorithm to propagate the dynamics of ions and virtual system variables that is accelerated by implementing a hybrid OpenMP/MPI parallelization approach to reduce the computing time associated with the evaluation of the forces and energies. The target applications of the framework are systems where the effects of NP surface charge and ion correlations typically lead to ion distributions that reach constant bulk value within a few nanometers of the NP surface such that a comprehensive study of ion densities near NP surfaces can be performed by including thousands of ions in a large simulation cell with reflective boundaries \citep{santos,messina1,boda}. Many synthetic and biological systems including oil-water emulsions, gold nanoparticles, and globular proteins exhibit this scenario, and ion distributions in these systems have been analyzed using $O(n^2)$ methods that are competitive with $O(n\log n)$ methods for these moderately-sized systems \citep{allen1,boda,santos,messina1,lue1,jso1}.
Further, attempts to ameliorate this scaling via the use of Ewald sums \citep{deserno1998mesh} or multigrid methods \citep{sagui2} would introduce more variables and parameters into the system making the assessment of the coupling of the ML-enabled parameter selection process with the simulation of ions and virtual system difficult. The use of such methods is thus avoided in this first study of developing ML-based enhancements for the treatment of polarizability effects in ionic soft-matter simulations; future work will include integrating this approach with fast Ewald solvers for incorporating long-range effects.

As we discuss later in the results section, the ML-enhanced dynamical optimization framework leads to an increase in both the efficiency and the stability of the associated MD simulations, while retaining the accuracy of the unautomated framework. This combination of ML and parallel computing in the context of nanoscale simulation of ions is the first of its kind and paves the way for developing online applications for web-based platforms like nanoHUB \citep{klimeck.nanohub}, where the user engages with the simulation software under limited interaction with the developer and/or domain expert. An application that simulates the self-assembly of ions near polarizable NPs by employing the unique features of this framework has been recently deployed on the nanoHUB cloud \citep{kadupitiya2018}. As is evident by the use of ML in numerous commercial platforms, scientific simulation workflow and software applications will increasingly employ an ML layer in the future. Understanding the integration of ML in scientific applications is thus critical; the work presented here contributes towards this goal. 

\section{Background and Related Work}\label{relatedwork}
\subsection{Model and the Energy Functional}
The problem of evaluating polarization effects in simulation of charged systems has been extensively explored by several research groups using different approaches \citep{marchi,santos,boda,tyagi,luijten.jcp2014,gan,jso1,all.atom2006,allen1}. Explicit simulation of solvent (environment) and NPs is possible \citep{all.atom2006} using advanced computational techniques such as fast multipole methods and local electrostatics algorithms \citep{rottler-maggs}; the role of solvent is also crucial in understanding properties of ion dynamics at the molecular scale \citep{beckstein2004not}. However, many phenomena can not be suitably investigated using fully atomistic models due to the prohibitively large number of degrees of freedom associated with such systems. This has led to the study of coarse-grained models that treat ions explicitly but replace the molecular structure of the solvent and the NP with continuous dielectric environments. Systems where the different material parts are adequately captured by piecewise-uniform dielectric permittivities (e.g. NP and solvent, protein and cellular medium) have attracted particular attention \citep{allen1,boda,santos,xu.icm2013,jso1,jso2,luijten.prl2014,tyagi}. For these model systems, solving for the induced charge density reduces the computational costs because the unknown induced charge density resides only on the two-dimensional interface (boundary) between the NP and the surrounding medium. We work with such a coarse-grained model. 

The dynamical optimization framework for extracting ion distributions from simulations of the coarse-grained model that treats the solvent and NP as dielectric continua is based on the true energy functional of the induced charge density introduced in \cite{jso1}:
\begin{equation}\begin{split}\label{eq:fnal}
\mathscr{F}[\omega]&=\frac{1}{2}\iint \rho_{\mathbf{r}}G_{\mathbf{r},\mathbf{r'}}
\left(\rho_{\mathbf{r'}}+\Omega_{\mathbf{r'}}[\omega]\right) \ud\mathbf{r'}\ud\mathbf{r}\\
&- \frac{1}{2}\iint \Omega_{\mathbf{r}}[\omega] G_{\mathbf{r},\mathbf{r'}}
\left(\omega_{\mathbf{r'}} - \Omega_{\mathbf{r'}}[\omega] \right) \ud\mathbf{r'}\ud\mathbf{r},
\end{split}\end{equation}
where $\rho$ and $\omega$ are the ion and induced charge densities respectively. The function $G(\mathbf{r},\mathbf{r'}) = |\mathbf{r} - \mathbf{r'}|^{-1}$ is the Green's function and $\Omega$ is given by 
\begin{equation}\label{eq:Omega}
\Omega_{\mathbf{r}}[\omega] = \nabla\cdot\left( \chi_{\mathbf{r}}\nabla\int 
\left( \rho_{\mathbf{r'}} + \omega_{\mathbf{r'}} \right) \ud\mathbf{r'}\right),
\end{equation}
where $\chi (\mathbf{r})$ is the dielectric susceptibility. $\chi(\mathbf{r})$ is related to the spatially-varying dielectric permittivity $\epsilon (\mathbf{r})$ via the relation $\epsilon = 1 + 4 \pi \chi$. The minimization of $\mathscr{F}[\omega]$ leads to the equation: 
\begin{equation}
\omega = \Omega.
\end{equation}
Solving this equation is equivalent to solving the Poisson equation; its solution produces the correct induced charge density \citep{jso1,jso2}. At its minimum, $\mathscr{F}[\omega]$ evaluates to the true electrostatic energy of the system. These features allow $\mathscr{F}[\omega]$ to be optimized dynamically as the ions move to their new positions in a simulation.

The functional $\mathscr{F}[\omega]$ can be transformed into a functional of only the surface (two-dimensional) induced charge density for the case of polarizable NP in a solvent where the NP and the solvent are modeled as materials of different, but uniform, permittivities \citep{jso1,jso2}. The discretized form of this transformed functional obtained by meshing the NP surface into $M$ finite elements is given as: \begin{align}\label{eq:dsfnal}
\mathscr{F}[\{\omega_{k}\}] &= \frac{1}{2}\sum_{i=1}^{N}\sum_{j\neq i}^{N}q_{i}
K^{^{^{\negthickspace\negthickspace\negmedspace\negthickspace\circ\circ}}}_{\mathbf{r}_{i},\mathbf{r}_{j}}q_{j}
+ \frac{1}{2}\sum_{i=1}^{N}\sum_{k=1}^{M}q_{i}
K^{^{^{\negthickspace\negthickspace\negmedspace\negthickspace\circ\bullet}}}_{\mathbf{r}_{i},\mathbf{s}_{k}}\omega_{k}a_{k}
\nonumber\\
&+ \frac{1}{2}\sum_{k=1}^{M}\sum_{l=1}^{M}\omega_{k}
K^{^{^{\negthickspace\negthickspace\negmedspace\negthickspace\bullet\bullet}}}_{\mathbf{s}_{k},\mathbf{s}_{l}}
\omega_{l}a_{k}a_{l},
\end{align}
where $\omega_k$, $\mathbf{s}_{k}$, and $a_k$ are, respectively, the induced charge, position vector, and area associated with the $k^{\textrm{th}}$ finite element. Here, $N$ is the total number of ions, and $q_{i}$ and $\mathbf{r}_{i}$ are the charge and position vector of the $i^{\textrm{th}}$ ion respectively. The terms
$K^{^{^{\negthickspace\negthickspace\negmedspace\negthickspace\circ\circ}}}$, $K^{^{^{\negthickspace\negthickspace\negmedspace\negthickspace\circ\bullet}}}$, 
and $K^{^{^{\negthickspace\negthickspace\negmedspace\negthickspace\bullet\bullet}}}$ in \eqref{eq:dsfnal}
are the effective potentials of interaction between two ions, between
an ion and an induced charge, and between two induced charges; explicit expressions of these functions can be found in the original papers \citep{jso1,jso2}. $\mathscr{F}[\{\omega_{k}\}]$ can be minimized on-the-fly using MD methods that treat the induced charges on the surface as dynamic variables; the details of this dynamical optimization framework are provided in Section \ref{cpmd}. We note that for the sake of brevity we have restricted the presentation of the above discretized form of the free-energy functional (Eq. \ref{eq:dsfnal}) to systems where the dynamics of ions is probed near a single dielectric interface. Extension of the approach to multiple dielectric surfaces corresponding to more complex systems (e.g., multiple ion-containing oil nanodroplets in water or electrolyte ions in water confined by material surfaces of different permittivities) is straightforward; results from these investigations have appeared in previous papers \citep{shen2017surface,jing2015}.

\subsection{Nanoscale Simulation of Ions near Polarizable Materials}\label{relatedwork.domain}
We review the techniques of computing the ion distributions in systems described by the coarse-grained model of ions near the dielectric interface separating NP and solvent \citep{boda, tyagi, allen1, luijten.jcp2014,jso1,jso2,jing2015}. Here, NP and solvent are characterized with different (but uniform) dielectric permittivities. In the light of the approach presented in Section 3, we focus on methods that are broadly applicable and are not limited by the choice of NP geometry or dielectric permittivity profile. 

We first outline the methods based on variational approaches to the problem of evaluating the polarization effects as these techniques are most closely related to the work presented here. In this approach, one transforms the original problem of solving the Poisson differential equation into an optimization problem. A variety of functionals employing various electrostatic quantities as field variables have been proposed to formulate the variational optimization problem \citep{jackson,marcus,felderhof,radke,karplus,allen1,attard,rottler-maggs,lipparini,VILLASENOR1992306,NAKANO1994181}. \cite{allen1} performed an explicit (static) optimization of a functional of the induced surface charge density \omegaS at each MD step to solve the Poisson equation and propagate ions. 
\cite{marchi} worked with a true energy functional of the polarization vector and implemented a dynamical optimization framework to propagate ion dynamics in parallel with the evaluation of polarization vector fields. However, the choice of the polarization vector as the variable field needed a three-dimensional specification leading to increased computational costs that can be avoided by choosing the induced charge density \omegaS as the variational field. 

Another class of methods for computing \omegaS transform the problem into a matrix formulation \citep{boda,tyagi,luijten.jcp2014}. The induced charge computation (ICC) methods \citep{boda} use matrix inversions to solve for $\omega(\mathbf{s})$. Matrix inversion operations involve $O(M^3)$ calculations where $M$ is the number of surface mesh elements. Techniques to improve upon this scaling have been subsequently developed \citep{tyagi}. Alternatively, iterative methods to solve the matrix equation have been proposed \citep{luijten.prl2014}. In particular, the generalized minimum residual method solves the matrix equation without explicitly constructing the inverse matrix and yields a converged result for \omegaS at each simulation timestep within 4 - 5 iterations \citep{luijten2015}. 

The evaluation of \omegaS in all the above approaches requires the ionic configuration to be static at each simulation step to guarantee the overall stability of the simulation. In Section \ref{cpmd}, we present the details of a recently developed dynamical optimization framework that enables the simultaneous (on-the-fly) updates of \omegaS and the ionic configuration in the same simulation step \citep{jso1,jso2}.

\subsection{Parameter Prediction using Machine Learning}\label{relatedwork.ml}
Machine Learning (ML) abstractions for parameter prediction and tuning have been extensively employed in the performance enhancement of bigdata or deep learning frameworks. \cite{denil2013predicting} used artificial neural network (ANN) and convolutional deep learning neural network (NN) to predict the parameters found in image classification tasks. The ANN was able to obtain an accuracy of 95$\%$. \cite{yigitbasi2013towards} employed ML-based auto-tuning for diverse MapReduce applications and cluster configurations in Hadoop framework. Their work showed that support vector regression (SVR) exhibits good accuracy while being computationally efficient for performance modeling of MapReduce applications.

Regression based prediction schemes have been employed in different domain areas \citep{eng2014predicting, kazemi2014one, cherkassky2004practical, chen2014short, balachandran2016adaptive, quan2014short, yadav2016selection}. \cite{eng2014predicting} used random forest regression algorithm to predict host tropism of influenza A virus proteins with an accuracy above $96\%$. Similarly, ensemble of regression trees were employed to perform face alignment for real-time applications (in one millisecond) by \cite{kazemi2014one}. SVR has been used for wind speed prediction by \cite{chen2014short}. ANN based regression has been studied by \cite{quan2014short} to yield short term load prediction of electrical power systems based on wind power forecasting.  \cite{yadav2016selection} have employed ANN based regression for forecasting solar radiation. 

In recent years, ML methods have been applied to enhance computational techniques aimed at understanding material phenomena; ML has been used to predict parameters, generate configurations in material simulations, and classify material properties \citep{glotzer2017,sam2017,fu2017,morningstar2017deep,melko2017,behler2016perspective,botu2015adaptive,long2015machine,ferguson2017machine,guo2018adaptive,ward2016general, kadupitiya2018machine,  kadupitiya2019machine3, kadupitige2019machine, fox2019learning, fox2019learning2}. 
\cite{glotzer2017} applied a simple feedforward ANN to discover interesting areas of parameter space corresponding to crystal formation in the self-assembly of colloidal building blocks. \cite{botu2015adaptive} employed kernel ridge regression (KRR) to accelerate the \emph{ab initio} MD method for nuclei-electron systems by learning the selection of probable configurations in MD simulations. \cite{fu2017} employed an ANN to select efficient updates for Monte Carlo simulations of classical Ising spin models. \cite{balachandran2016adaptive} have used SVR to create an adaptive ML model to aid the design of new materials with desired elastic properties and enhanced long-term performance  using minimum number of iterations. 

These explorations have inspired us to use ML to design an adaptive MD-based dynamical optimization framework that updates the simulation timestep and auto-tunes the virtual parameters characterizing the dynamics of ions near polarizable NPs to yield a more stable and efficient simulation. Related work in the area of adapting timestep in a simulation has involved using analytical approaches to multiple timestep integration \citep{martinez2014,tuckerman1992reversible}. Recent work has also focused on adaptive ensemble simulations to enhance the computational efficiency of biomolecular simulations \citep{kasson2018adaptive}. We also note the development of auto-tuning technology for high-performance computing applications to reduce execution time and enhance programmer productivity \citep{whaley1998automatically}. Here, auto-tuning relates to the automatic generation of a search space of possible kernels for a computational task to identify the best possible kernel, with recent work involving the use of ML-based approaches for identifying the search space \citep{balaprakash2018autotuning}. In Section ~\ref{ml.framework}, we describe the results of our experiments with different regression-based ML models to identify and tune optimal simulation parameters in MD simulations based on the dynamical optimization framework.

\section{Dynamical Optimization Framework for Simulating Ions near Polarizable NPs}\label{cpmd}
In this section, we provide the details of the dynamical optimization framework for simulating ions in the presence of polarizable NPs. This framework uses Car-Parrinello molecular dynamics (CPMD) technique \citep{car-parrinello,fois} to dynamically optimize an energy functional of the polarization charge density which results in the propagation of the ionic configuration in tandem with an accurate update of the polarization charges \citep{jso1,jso2}. These details will help clarify the use of the ML-based enhancement strategies outlined in Section \ref{ml.framework}. 
\subsection{Extended Lagrangian, Equations of Motion, and CPMD Simulation}
To implement the dynamical optimization of $\mathscr{F}[\{\omega_{k}\}]$, the induced charges $\{\omega_{k}\}$ are treated as dynamic virtual variables. A fictitious kinetic energy is associated with this virtual system:
\beq
\mathscr{K} = \sum_{k=1}^{M}\frac{1}{2}\mu_k \dot{\omega}^{2}_{k},
\eeq
where 
$\mu_k$ is the mass of the $k^{\mathrm{th}}$ virtual variable $\omega_k$, and $M$ is the number of mesh points discretizing the NP surface.
The extended Lagrangian $\mathcal{L}$ with \functional as its electrostatic potential energy is constructed by including $\mathscr{K}$ as an additional term:
\begin{equation}\label{eq:lag}
\mathcal{L} =  \mathscr{K} + \sum_{i=1}^{N} \frac{1}{2}m_{i}\dot{\mathbf{r}}^{2}_{i} - \mathscr{F}[\{\omega_{k}\}]-\mathscr{H}[\{\mathbf{r}_i\}].
\end{equation}
In \eqref{eq:lag}, the second term is the usual total kinetic energy associated with $N$ ions (physical system), with $m_i$ being the mass of the $i^{\mathrm{th}}$ ion. 
The final term contains a set of Lennard-Jones potentials to model the ion-ion and ion-NP steric interactions. Note that \functional$\equiv\mathscr{F}[\{\omega_k\},\{\mathbf{r}_i\}]$ is also a function of the set of ion positions $\{\mathbf{r}_i\}$.

The Lagrangian $\mathcal{L}$ yields the following Euler-Lagrange equations of motion:
\begin{align}
\begin{split}\label{eq:eomreal}
m_{i}\ddot{\mathbf{r}}_{i} &= -\nabla_{\mathbf{r}_{i}}\mathscr{F}[\{\omega_{k}\},\{\mathbf{r}_{i}\}],
\end{split}
\\
\begin{split}\label{eq:eomfake}
\mu_{k}\ddot{\omega}_{k} &= -\nabla_{\omega_{k}}\mathscr{F}[\{\omega_{k}\},\{\mathbf{r}_{i}\}],
\end{split}
\end{align}
for the $i^{\textrm{th}}$ ion and the $k^{\textrm{th}}$ induced charge, respectively. These equations are used to evolve the induced charge configuration \emph{on the fly} using the CPMD method. Following \eqref{eq:eomreal}, each ion is moved by the force $-\nabla_{\mathbf{r}_{i}}\mathscr{F}[\{\omega_{k}\},\mathbf{r}_{i}]$ in a timestep $\Delta$ during which each induced charge is updated via the \emph{force} 
$-\nabla_{\omega_{k}}\mathscr{F}[\{\omega_{k}\},\mathbf{r}_{i}]$ following \eqref{eq:eomfake}.
To simulate the behavior of the ions at temperature $T$, the extended system of ions and virtual variables is coupled to a set of Nos$\acute{\textrm{e}}$-Hoover thermostats (this coupling modifies the equations of motion \eqref{eq:eomreal} and \eqref{eq:eomfake} similar to a canonical MD routine). This two-temperature approach is a standard feature of CPMD \citep{sprik,blochl-parrinello,fois}. The ions couple to a thermostat at temperature $T$, while the virtual system is coupled to one at $T_v$. 

Velocity-Verlet algorithm is used to generate the dynamics of the extended system. The dynamics is associated with a conserved quantity, the total energy of the extended system:  
\begin{equation}\label{eq:exte}
\mathscr{E} = \sum_{i=1}^{N} \frac{1}{2}m_{i}\dot{\mathbf{r}}^{2}_{i} + \mathscr{K} + \mathscr{F}[\{\omega_{k}\}] + \mathscr{H}[\mathbf{r}_i] + \mathcal{T} + \mathcal{T}_v.
\end{equation}
Here, $\mathcal{T}$ and $\mathcal{T}_v$ are the energy terms associated with the thermostats controlling the temperature of the physical and virtual systems respectively. The extended energy and $\mathscr{K}$ are monitored at periodic intervals during a CPMD simulation to assess the stability and accuracy of the simulation.

Virtual masses $\mu_k$ are chosen to be proportional to the areas of the mesh points. The value of the proportionality constant $\mu$ depends on the attributes of the physical system (e.g., NP charge, dielectric profile, ion valencies) as well as the simulation timestep $\Delta$. The parameters $T_v$ and $\mu$ are optimized to ensure the stability and accuracy of the simulation (see Section \ref{cpmd.feature}); further technical details of the method can be found in \cite{jso1}.

\subsection{OpenMP/MPI Hybrid Parallelization}\label{pc.cpmd}
A system with $N$ ions near an unpolarizable NP effectively translates into a system with $M$ additional dynamical variables in the case of a polarizable NP within the dynamical optimization framework. Due to the long-range nature of the electrostatic interactions, the associated computational costs scale roughly as $O((N+M)^2)$, with a prefactor that can be large owing to the complexity of the terms involved in the expressions for forces derived from $\mathscr{F}[\{\omega_k\}]$. Indeed, performance profiling report generated using Performance Counters for Linux (PERF) showed that the sequential program spends the largest amount of computation time ($64.33\%$ of the total) calculating the forces between the ions for each step of the simulation. To reduce the computing time associated with the evaluation of these forces and enhance the performance of the simulation framework, a hybrid OpenMP/MPI parallel programming model is adopted. The hybrid model has advantages over pure MPI or pure OpenMP, when cache performance is taken into consideration. This strategy provides non-uniform memory access (NUMA) traffic and inter-node communication \citep{rabenseifner2009hybrid} to support maximum access locality and minimum number of cache misses. We note that the simulations associated with the dynamical optimization framework presented in the previous publications employed the OpenMP (shared memory) parallelization model, and consequently, were limited in their application scope \citep{jso1,jso2,jing2015}.

\begin{figure}[th]
\centerline{\includegraphics[scale=0.85]{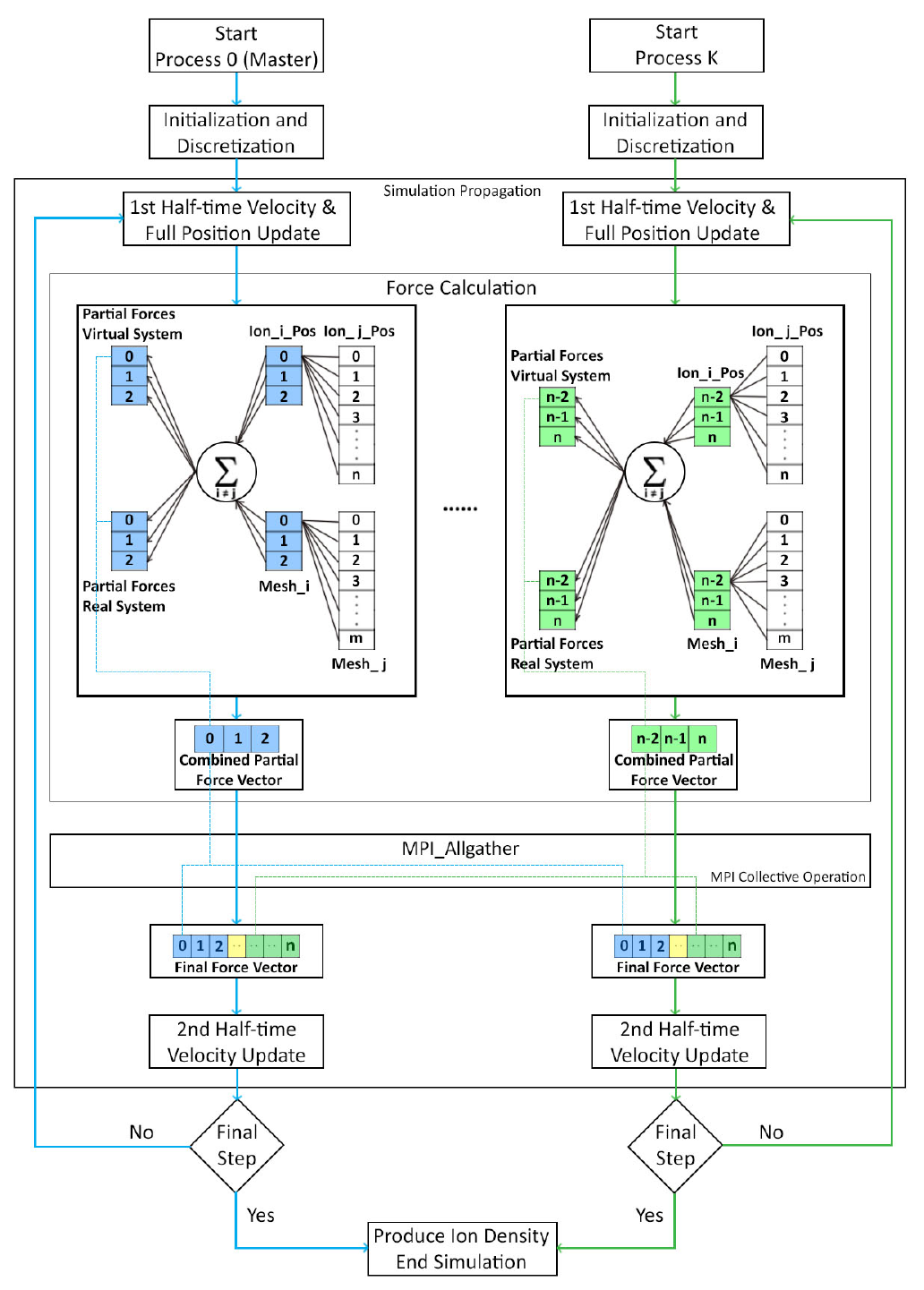}}
\caption
{\label{mpi.overview}
Hybrid model (employed inside the Force Calculation block) with distributed and shared memory parallelization techniques implemented using MPI and OpenMP.
}
\end{figure}

The hybrid masteronly model is implemented by combining the distributed memory MPI approach and the shared memory OpenMP approach \citep{rabenseifner2009hybrid}, and is applied to the force and energy calculation subroutines in the dynamical optimization framework. The model uses one MPI process per node and OpenMP on the cores of the node, with no MPI calls inside the parallel regions. The domain decomposition is enabled under a two-level mechanism. On the MPI level, a coarse-grained domain decomposition is performed using boundary conditions as explained in Fig. \ref{mpi.overview}. The second level of domain decomposition is achieved through OpenMP loop level parallelization inside each MPI process. 

\subsection{Key Framework Features}\label{cpmd.feature}
We identify two key features of the dynamical optimization framework that encode the accuracy and stability of the simulation, and guide the process of designing the ML-based enhancement strategies presented in Section \ref{ml.framework}.
The first key feature is the conservation of the extended energy $\mathscr{E}$, given by Eq. \eqref{eq:exte}, and the approximate conservation of the energy of the physical system that is captured by demanding that the kinetic energy of the virtual system nearly vanishes:
\beq\label{eq:key.condition}
\mathscr{K} \approx 0.
\eeq
In other words, the framework ensures that the physical system remains unaffected as much as possible by the presence of the virtual system. 

\begin{figure}[t]
\centerline{\includegraphics[scale=0.69]{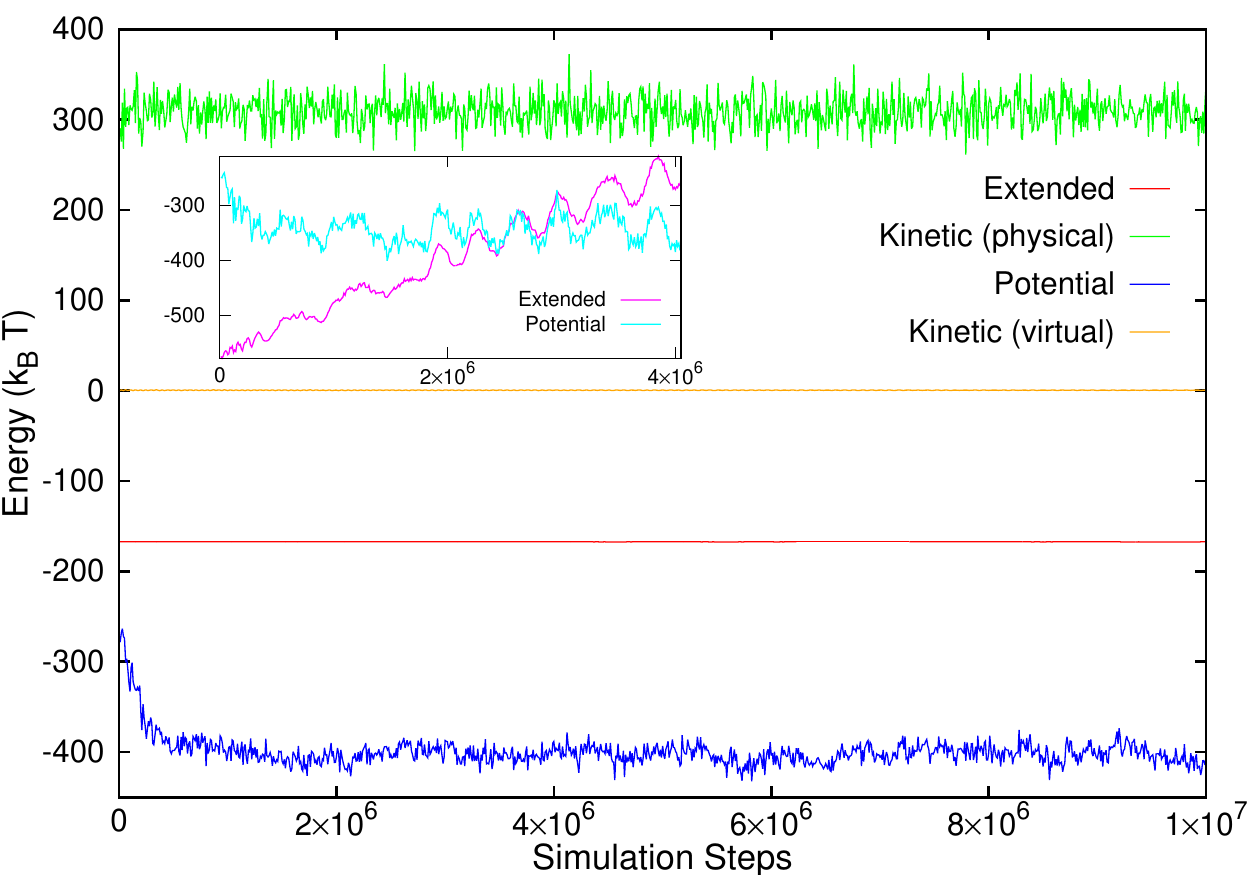}}
\caption
{\label{energy.method}
Energy profiles of 206 ions near a polarizable NP (whose surface is meshed with 1082 grid points). (Outset) Data is shown for $\approx 10$ nanoseconds of simulated physical time (10 million simulation steps). Conservation of the extended (total) energy and nearly vanishing ($\approx 0$) virtual kinetic energy enabled by the selection of ``good'' virtual parameters ($\mu =15$, $T_v =0.001$) highlight the first key feature of the dynamical optimization framework. (Inset) Energy profiles for the same system when a set of ``bad'' virtual parameters are selected ($\mu =6$, $T_v =0.05$). This selection produces unphysical dynamics of the ionic system with the extended energy not getting conserved and the potential energy exhibiting artificial oscillations.}

\end{figure}

The energy profiles of a typical, successful CPMD simulation of ions near a polarizable, spherical NP at room temperature are shown in Fig.~\ref{energy.method}: the extended energy $\mathscr{E}$ is constant and the total virtual kinetic energy $\mathscr{K}$ stays stable and close to 0 throughout the entire simulation (for $\approx 10$ ns). In practice, this feature is incorporated in the simulation by appropriately choosing values of simulation timestep $\Delta$, virtual variable mass $\mu$, and the virtual system temperature $T_v \ll T$ ($T_v \approx 0$). These parameters are selected to control large, abrupt rise in the kinetic energy associated with the virtual system as the simulation progresses.

This feature is encoded in the quantity $R$ which measures the ratio of the fluctuations in $\mathscr{E}$ and the fluctuations in the kinetic energy of the physical system. $R$ determines the stability and the accuracy of the simulation. For good energy conservation (constant $\mathscr{E}$), it is demanded that the simulations satisfy the condition $R < 0.05$ as noted in the literature \citep{marchi}. The latter inequality implicitly satisfies the requirement that $\mathscr{K}$ is kept close to the value dictated by the low temperature $T_v$.

\begin{figure}[t]
\centerline{\includegraphics[scale=0.84]{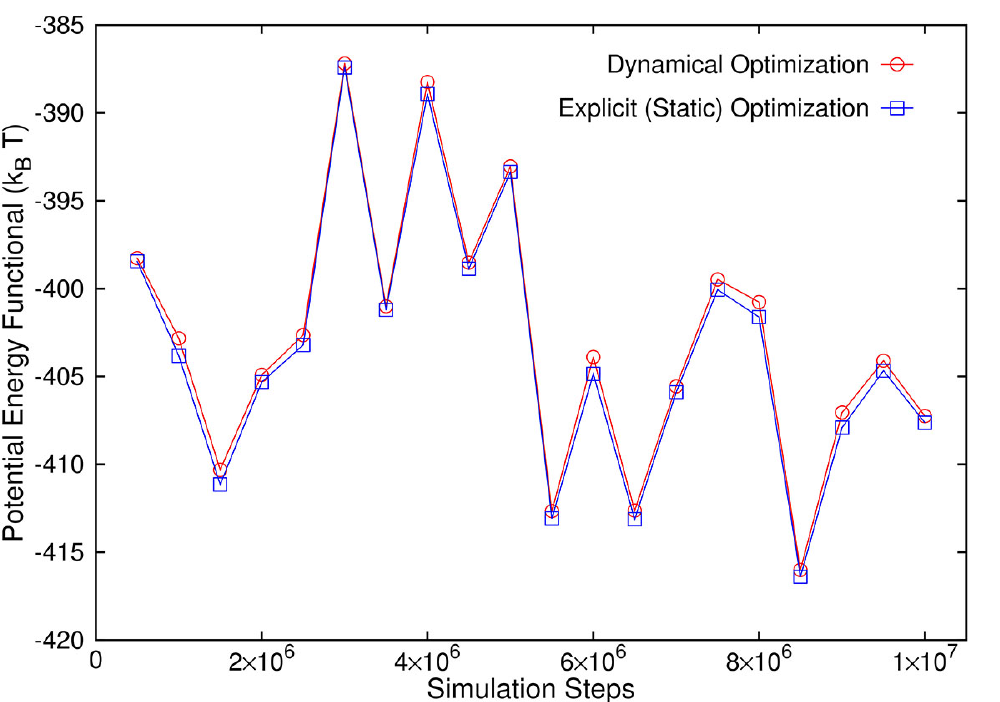}}
\caption
{\label{tracking.method}
Comparison of the functional optimized dynamically (circles) and the functional optimized at regular intervals keeping the ionic configuration static during the optimization process (squares). Results are shown for the same system as in Fig. \ref{energy.method}. The matching of the two functionals illustrates the second key feature of the framework: the accurate tracking of the induced charge density.
}
\end{figure}

The second important feature considers the effectiveness of the framework to reproduce the induced charge distribution accurately at each simulation step. At regular intervals during the course of the simulation, the ion coordinates and induced charge densities on the NP surface are stored. Then, an ordinary (static) minimization of the functional $\mathscr{F}$ is carried out to explicitly determine the (numerically) exact induced charge density. The tracking of the induced charge density distributions on the NP surface can be assessed by evaluating the matching of $\mathscr{F}$ optimized on the fly with the electrostatic energy value obtained by optimizing the functional explicitly (Fig. \ref{tracking.method}). This functional matching is the second key feature. In practice, we compute the functional deviation, $f_d$, which measures the average difference between the dynamically optimized functional $\mathscr{F}$ and the energy functional obtained via direct (static) minimization. To pass the test of stability and accuracy, we enforce $|f_d| < 1 \%$. 

$R$ and $f_d$ are central to the success of the simulations based on this framework and determine the associated ``good'' virtual system parameters $\mu$ and $T_v$. In general, higher $\mu$ leads to better energy conservation and lower $R$, while lower $T_v$ keeps the virtual system from excessive heating and generates lower $f_d$. Having just these two features  
biases the prediction of $\mu$ ($T_v$) towards higher (lower) values for a system of ions characterized with a generic input parameter pattern. Very high values of $\mu$ and/or very low values of $T_v$ can affect the overall stability of the simulation as the virtual system can be prohibitively slow (due to the ``heavy'' virtual masses and ``cooler'' associated temperature) to react to the evolving ionic configuration, resulting in inaccurate induced charge updates. An experienced domain expert would typically avoid these choices. To enhance the stability of the simulation and bias the selection of the virtual parameters towards those picked by a domain expert, another quantity, $R_v$ is introduced. $R_v$ is the ratio of the fluctuations in $\mathscr{E}$ and the fluctuations in the kinetic energy of the virtual system $\mathscr{K}$. Lower $R_v$ implies that fluctuations in $\mathscr{K}$ (that can arise from lower $\mu$ or higher $T_v$) are sufficiently strong to endow the virtual system with the necessary dynamics to adapt to the evolving ionic system. Unlike $R$ and $f_d$ that exhibit universal bounds informed by the physical dynamics of ions, the bound on $R_v$ depends on the set of systems investigated, and is informed largely by past domain experience. For counterion-only systems used as training set for the ML-based methods, $R_v < 0.15$ is enforced. For systems characterized with electrolytes that are expected to exhibit a greater number of ions (with both positive and negative valencies), $R_v$ can assume larger values depending on the salt concentration. 

Quantities $R$, $f_d$, and $R_v$ determine the choice of the optimal virtual parameters. However, these quantities and the success of the simulation depend critically on another important parameter: the simulation timestep $\Delta$. The simulation timestep in the CPMD-based dynamical optimization framework depends on both the physical and the virtual system parameters, the latter being unknown \emph{a priori}. Conversely, the optimal values of virtual parameters ($\mu$, $T_v$) that ensure a long-time stable simulation are dependent on $\Delta$. This complicates the process of choosing a reliable, yet efficient, value for $\Delta$, $\mu$, and $T_v$ and one typically chooses a conservative $\Delta$ that is smaller than the value used in conventional MD simulations ($\Delta \approx 1 - 5$ femtoseconds for an MD simulation of monovalent electrolyte ions in water at room temperature).

\section{ML-based Enhancement of the Dynamical Optimization Framework}\label{ml.framework}
We now describe an ML-based procedure to increase the efficiency and improve the stability of the dynamical optimization framework while retaining the simulation accuracy. We present an ML technique that uses the aforementioned key features encoded in quantities $R$, $f_d$, and $R_v$ to enhance the performance of the dynamical optimization framework by 1) predicting and auto-tuning the optimal virtual parameters $\mu$ and $T_v$, and 2) adapting the timestep to the largest allowable value during the simulation. The ML technique is combined with OpenMP/MPI Hybrid parallel programming model, described in Section \ref{cpmd}, to carry out the simulation.

\begin{figure}[th]
\centerline{\includegraphics[scale=0.55]{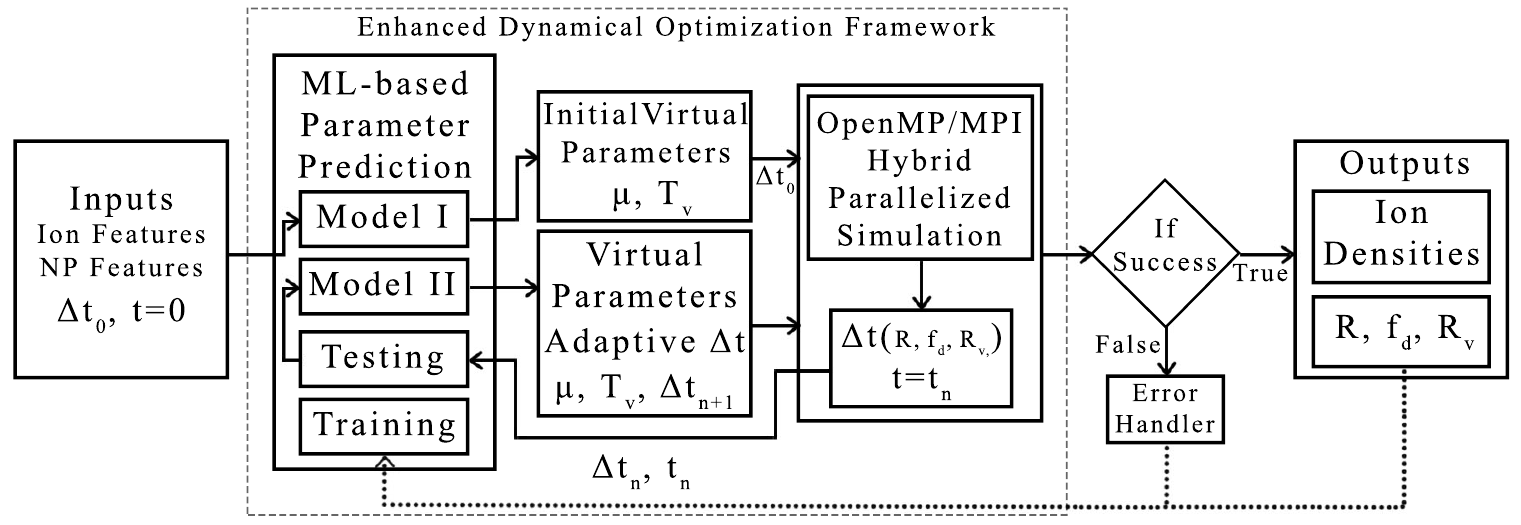}}
\caption
{\label{fig.overview}
System overview of the ML-enhanced dynamical optimization framework.
}
\end{figure}

Figure \ref{fig.overview} shows the overview of the enhanced framework. ML-based parameter prediction was implemented using two ML models (ML model I and II). First, the ion and NP model attributes, as well as the initial timestep $\Delta = \Delta t_0$, were fed to ML model I to predict the initial virtual system parameters $\mu$ and $T_v$. The predicted parameters and $\Delta t_0$ were used to start the simulation that was parallelized using the OpenMP/MPI hybrid programming model. At intermediate times $t_n$ during the simulation, ML model II was used to predict the new timestep $\Delta t_{n+1}$ and associated virtual parameters that continue the simulation for the subsequent time block $(t_n, t_{n+1})$. The ion distributions near the polarizable NP were sampled during the simulation run and the ion densities were stored as simulation output. ML model II also checked if the simulation was successful up to $t_n$ before dynamically tuning the parameters for the next iteration. The program aborted and called the error handler to display appropriate error messages if the simulation failed due to the imposed $R$, $f_d$, and $R_v$ criteria. In addition, during the simulation, the quantities $R$, $f_d$, and $R_v$ were computed and saved as output for retraining both ML models after a set number of simulation runs were executed. For every 1000 new simulation runs, both models were retrained. 

\begin{figure}[t]
\centerline{\includegraphics[scale=0.8]{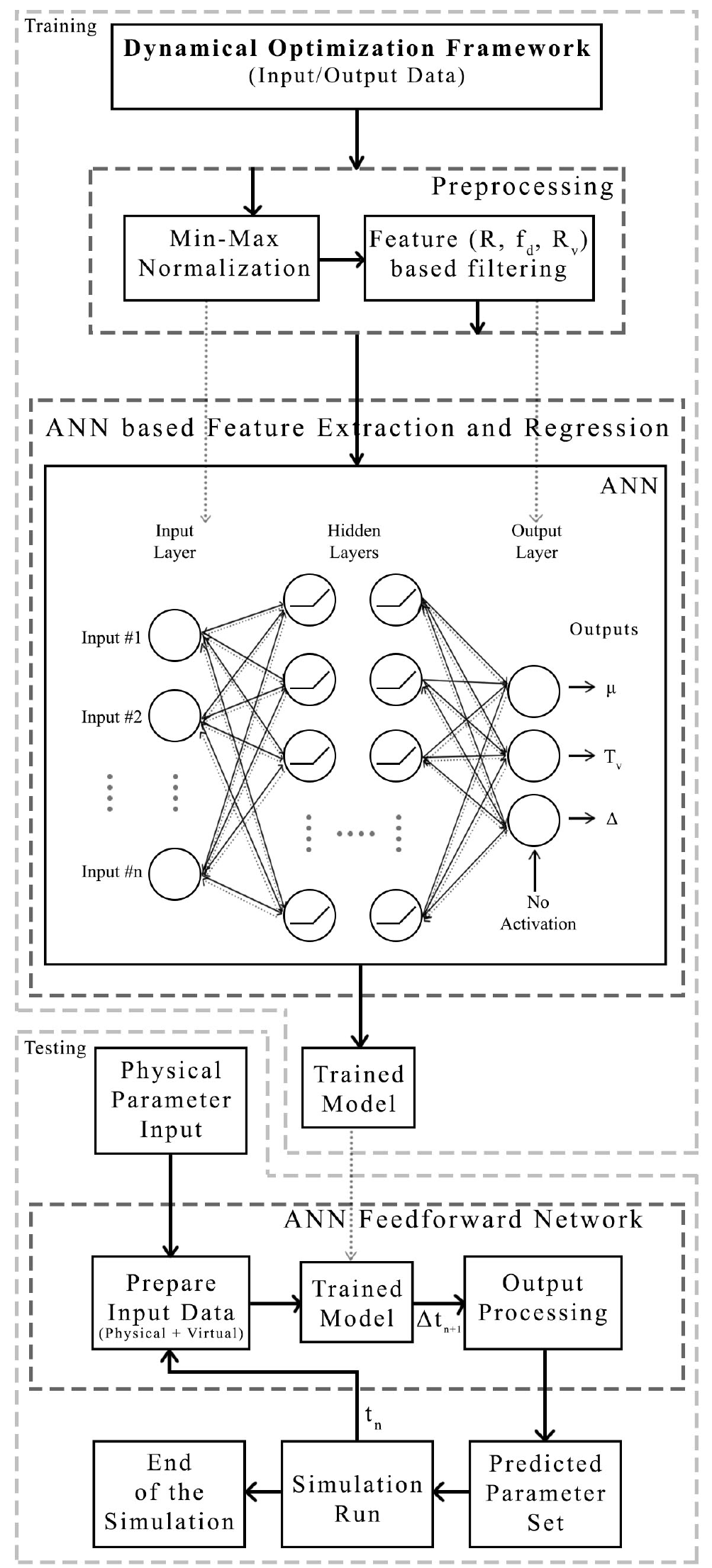}}
\caption
{\label{ml.overview}
ML procedure (model II) for determining the parameters of the virtual system and the adaptive timestep used in the dynamical optimization framework.
}
\end{figure}

After reviewing and experimenting with many ML techniques for parameter tuning and prediction including polynomial regression, support vector regression (SVR), decision tree regression, and random forest regression (Section \ref{relatedwork.ml} and Section \ref{results.ml.tech}), the artificial neural network (ANN) was adopted for enhancing the dynamical optimization framework. Figure \ref{ml.overview} shows the details of the ANN-based ML model II employed to predict the virtual system parameters and the adaptive timestep for simulations based on the framework; the ANN was trained to select the largest allowable $\Delta$ that satisfies the tests of stability and accuracy encoded in the ML features. ML model I exhibits a similar process but is trained without the time and timestep parameters.
The data preparation and preprocessing techniques, feature extraction and regression techniques as well as their validation for both models are discussed below. 

\subsection{Data Preparation and Preprocessing}
Counterion-only systems (no added electrolyte) were considered for generating the training set. The polarization effects are expected to be strongest in these systems as added electrolyte screens the ion-NP electrostatic interactions. Further, counterion-only systems are relatively smaller and enable a broader exploration of parameter space to train the ML models. Interestingly, as we discuss later, results from counterion-only systems were employed to successfully extract and infer ionic distributions associated with electrolyte systems for up to $O(0.1)$ M concentration, exhibiting the transfer learning aspects of the ML-based procedure employed. 

Prior domain experience and backward elimination using the adjusted R squared was considered for creating the training data set. Using this process, 5 input parameters that significantly affect the polarization charges on the NP surface were identified: NP permittivity $\epsilon_{\textrm{NP}}$, solvent permittivity $\epsilon_{\textrm{S}}$, NP charge $Q$ (in units of electronic charge $|e|$), counterion valency $v$, and NP mesh size $M$. While the temperature of the physical system and the size of the NP and ions affect the polarization charges and associated ionic distributions, in this initial study, these were considered fixed; NP diameter was taken to be $\approx 7.5$ times the ion diameter ($0.357$ nanometers), and temperature was fixed at 298 K. 
Despite the aforementioned potential for transfer learning associated with the ML procedure, additional parameters such as salt concentration and co-ion attributes should be included in the training set to predict the optimal parameters associated with the simulation of electrolyte systems with higher accuracy; future work will explore the training with these additional input parameters. Further, to enable accurate prediction of virtual parameters and generation of stable ion dynamics near NPs of different shapes (e.g., discs, rods) or NPs present in media characterized with a more complex permittivity profile (e.g., NPs stabilized near an oil-water interface), the training data must include NP shape characteristics and the relevant permittivity profile characterizing the system.
Virtual parameter mass $\mu$ and virtual system temperature $T_v$ were selected as the output parameters. Few discrete values for each of the input/output parameters were experimented with and swept over to create and run 13,600 simulations for training the ML model I. The range for different ionic system parameters was selected based on physically meaningful and experimentally-relevant values: $\epsilon_{\textrm{NP}} \in (2, 160)$;  $\epsilon_{s} \in (2, 160)$; $Q \in (-20, -100)$; $v \in 1,2,3$. For the mesh size and the virtual system parameters, the range was chosen based on previous trial and error procedure: $M\in (132, 1692)$, $\mu$ was swept from 1 to 40 using random discrete values to cover the range, and $T_v$ was swept from 0.001 to 0.005. All simulations were performed for $\approx 1$ nanoseconds.

To support on-the-fly tuning of $\Delta$ and associated selection of $\mu, T_v$ during the simulation, ML model II was trained with two additional parameters. Simulation time $t \equiv t_n$ and timestep $\Delta \equiv \Delta t_{n\to n+1}$ were added as input and output parameters respectively to the system parameters explored in ML model I. 20 discrete simulation time values $t_n \approx 0.1, 0.2, \ldots, 2$ ns, and 4 discrete timestep values $\Delta = 0.001, 0.002, 0.003, 0.004$, were swept to generate 54,400 simulation configurations.
In our current approach, the time interval (and related frequency) between updates of the simulation timestep and virtual parameters is $\delta = 0.1$ nanoseconds. This update frequency is largely limited presently by the computing costs associated with the size of the dataset needed to compute reliable estimates of $R, R_v$, and $f_d$ at $\delta$ time intervals. A higher frequency (lower $\delta$) will require a larger training dataset to be generated, adding to the computational costs. Further, for good stability profile, maximum update frequency should be smaller than characteristic timescales of fluctuations in the induced charge density (indicative of changes in the virtual system energy) in response to the dynamics of ions near polarizable surfaces.

As described in Section \ref{cpmd.feature}, $R$, $f_d$, and $R_v$ encode the key features of energy conservation and accurate tracking of the induced charges that measure the success of the dynamical optimization framework. Acceptable threshold values were identified for $R$, $f_d$, and $R_v$ as 0.05, 1$\%$, and 0.15 respectively for the range of systems included in the training set. These quantities were treated as output features to filter the datasets to only keep the input parameter configurations resulting in successful simulation runs. From the data set for initial parameter prediction, 4530 input/output configurations were selected as successful. From the data set for adaptive timestep prediction, 15640 input/output configurations were selected based on the same $R$, $f_d$, and $R_v$ criteria. Each of these datasets were separated as training and testing using a ratio of 0.7:0.3. Min$-$max normalization filter was applied to normalize the input data in the data preprocessing stage. 

\subsection{Feature Extraction and Regression}
The ANN algorithm with two hidden layers (Fig. \ref{ml.overview}) was implemented in Python for regression of two continuous variables in ML model I, and for regression of three continuous variables in ML model II. In both models, outputs of the hidden layers were wrapped with the \emph{relu} function; the latter was found to converge faster compared to the sigmoid function. No wrapping functions were used in the output layers of the algorithm. 

By performing a grid search, hyper-parameters such as the number of first hidden layer units, second hidden layer units, batch size, and number of epochs were optimized to 13, 8, 25, and 150 respectively. Adam optimizer was used as the back propagation algorithm. The weights in the hidden layers and in the output layer were initialized to random values using a normal distribution at the beginning. The mean square loss function was used for error calculation in both ML models. We use dropout regularization mechanism (where hidden layer neurons are turned off at a dropout rate of 0.2) to prevent overfitting which was evaluated using the k-fold cross-validation technique.
ANN implementation, training and testing was programmed with the aid of keras and sklearn ML libraries \citep{chollet2015keras,buitinck2013api, abadi2016tensorflow}.

\section{Results and Discussion}\label{results}
\subsection{Initial Virtual Parameter Prediction}\label{results.ml.tech}
Several regression models were implemented and tested to predict the initial virtual parameters $\mu$ and $T_v$. These models were tested on 1359 input parameter sets comprising of values within the range for which the models were trained for. Additionally, the models were tested on 450 completely random input sets of parameters, which included some selections beyond the range of training dataset. Table \ref{tab.ml.models} shows the success rate and mean square error for testing data sets as well as random input data sets. Success rate was calculated based on $R$, $f_d$, and $R_v$. Reported mean square error (MSE) values are calculated using k-fold cross-validation techniques with $\mathrm{k} = 10$. ANN based regression model predicted the initial virtual system parameters correctly with $94.3\%$ success rate (MSE of 0.56), outperforming other non-linear regression models as evident from Table \ref{tab.ml.models}. ANN based regression model was also able to achieve a success rate of $87.2\%$ on completely random input parameters. This ANN based regression model was adopted as ML model I.

\begin{table}[htbp]
\caption{Comparison of regression models for the prediction of initial virtual parameters.}
\begin{center}
\begin{tabular}{|c|c|c|c|}
\hline
\multicolumn{1}{|c|}{\textbf{Model}}&\multicolumn{2}{|c|}{\textbf{Testing Sets}}&\multicolumn{1}{|c|}{\textbf{Random Sets}} \\
\cline{1-4}
 & \emph{Success \%} & \emph{MSE} & \emph{Success \%} \\
\hline
Polynomial & 45.7& 12.25& 16.5\\
\hline
Support Vector & 78.3& 4.56& 58.4\\
\hline
Decision Tree & 70.4& 6.93& 48.1 \\
\hline
Random Forest & 75.3& 4.88& 55.1\\
\hline
ANN based & 94.3& 0.56& 87.2\\

\hline
\end{tabular}
\label{tab.ml.models}
\end{center}
\end{table}

Table \ref{tab.ml} shows the predicted $\mu$ and $T_v$ for selected systems along with the quantities $R$, $f_d$, and $R_v$ that characterize the key features of the framework: energy conservation and tracking of the induced charges. The predicted $\mu$ and $T_v$ values produced stable and accurate dynamics of ions near polarizable NP as evidenced by the values of $R$, $f_d$, and $R_v$ that lie within the allowed ranges ($R<0.05$, $|f_d| < 1 \%$, $R_v<0.15$). 

\begin{table}[htbp]
\caption{Predicted parameters by ML model I, and simulation accuracy and stability.}
\begin{center}
\begin{tabular}{|c|c|c|c|c|c|}
\hline
\multicolumn{1}{|c|}{\textbf{Inputs}}&\multicolumn{2}{|c|}{\textbf{Prediction}}&\multicolumn{3}{|c|}{\textbf{Results}} \\
\cline{1-6}
$\epsilon_{\textrm{NP}}$, $\epsilon_\textrm{S}$, $Q, v$ & \textbf{$\mu$ } & \textbf{\textit{$T_v$ }} & \textbf{\textit{$R$}}& \textbf{\textit{$f_d$}}& \textbf{\textit{$R_v$}} \\
\hline
2, 10, -60, 1& 9& 0.001& 0.001& -0.01 & 0.12 \\
\hline
2, 78.5, -30, 3& 7& 0.002& 0.003&  -0.6 & 0.13\\
\hline
50, 78.5, -60, 2& 18& 0.001& 0.001& -0.4 &  0.06 \\
\hline
80, 160, -90, 3& 30& 0.002& 0.002& -0.7&  0.09 \\
\hline
100, 120, 30, 2& 36& 0.005& 0.002& -0.1& 0.10\\
\hline
5, 71, -24, 2& 42& 0.001& 0.007& -0.6 & 0.13  \\
\hline
44, 37, -114, 1& 38& 0.006& 0.002& -0.11 & 0.11 \\
\hline
30, 35, -108, 3& 43& 0.007& 0.002& -0.12 & 0.05 \\
\hline
15, 78, -102, 1& 17& 0.025& 0.005& -0.27& 0.11 \\
\hline
\end{tabular}
\label{tab.ml}
\end{center}
\end{table}

We note that when the ANN was trained utilizing only $R$ and $f_d$ quantities, higher $\mu$ and lower $T_v$ values were predicted as expected from the arguments presented in Section \ref{cpmd.feature}. These virtual parameter choices are not optimal for the stability of the simulation and will not be picked by an experienced, domain expert. Inclusion of $R_v$ as another feature for training the model enabled the ANN to predict the virtual system parameters that were likely to enhance the stability of the simulation and be selected by an expert.  

\subsection{Auto-tuning CPMD Simulation Parameters}
Similar to ML model I, ML model II employed the ANN based regression model trained with two additional parameters: simulation timestep $\Delta$ and time $t$. This model was trained to infer the largest allowed $\Delta$ and auto-update the associated optimal virtual parameters $\mu, T_v$ at $t_n$ for the simulation during the interval ($t_n, t_{n+1}$) based on the ML output features $R, f_d$, and $R_v$ in the time interval $(t_n, t_{n+1})$ from the training data. 

\begin{figure}[th]
\centerline{\includegraphics[scale=0.84]{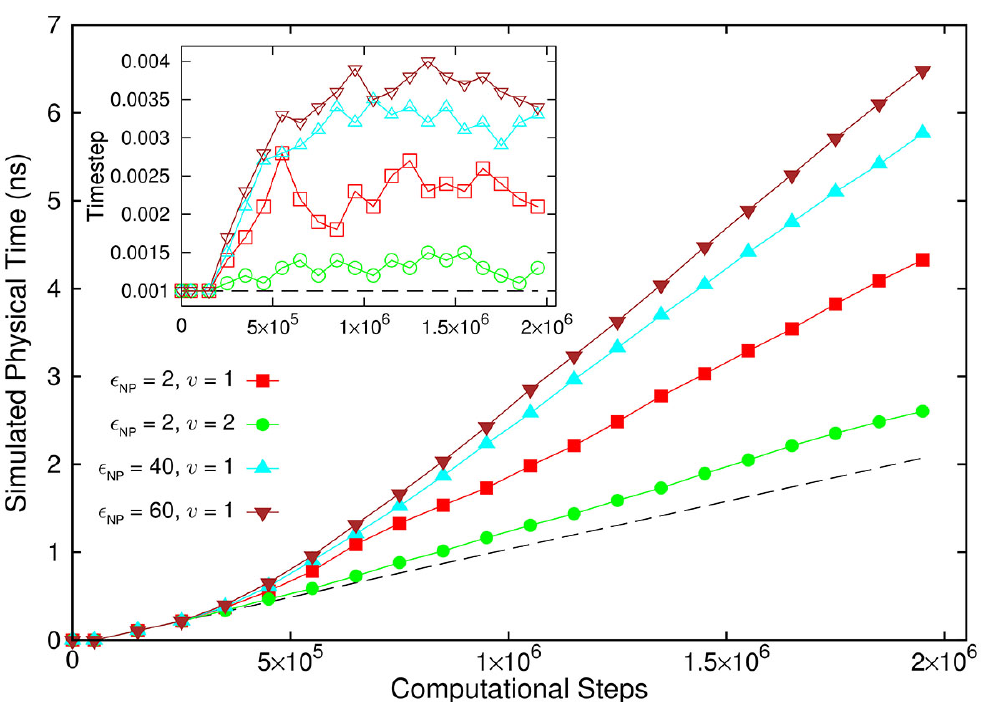}}
\caption
{\label{ml.speedup.comparison}
Simulated physical time $t$ associated with the dynamics of ions as a function of computational steps $S$ for a system of counterions near a polarizable NP. The legend denotes NP permittivity and ion valency ($\epsilon_{\textrm{NP}}; v$). Symbols are the results from using ML-enabled tuning of simulation timestep $\Delta$ and the dashed line is the result for the non-adaptive case. (Inset) $\Delta$ in units of $\approx 1.09$ picoseconds. Open symbols correspond to the systems denoted with the closed symbols in the legend; black dashed line denotes $\Delta = 0.001$ associated with the non-adaptive case. $\Delta$ values represent the average timestep over a period of $t_n \to t_{n+1}$.
}
\end{figure}

Figure \ref{ml.speedup.comparison} illustrates the computational gains resulting from the auto-tuning of $\Delta$ using ML model II for systems of counterions near a nanoparticle (NP). NPs of different dielectric permittivity ($2, 40$ and $60$) in water (dielectric permittivity $78.5$) are considered. The effect of varying ion valency ($v = 1, 2$) is also probed. Other input parameters are NP charge $Q = -100$, NP mesh size $M = 1272$, and $\epsilon_{\textrm{S}} = 78.5$. Symbols indicate the gains associated with the enhanced framework with adaptive timestep. The dashed line is the result from the non-adaptive simulation with static timestep for $\epsilon_{\textrm{NP}} = 2, v = 1$ case (other systems also yield the same result when non-adaptive model is used). Compared to the simulation with non-adaptive $\Delta$ (performed using only ML model I), the auto-tuning of $\Delta$ extended the simulation of the ionic system to a longer physical time for the same number of simulation steps; a speedup of $\approx 1.25 - 3$ was observed depending on the system configuration, also see Table \ref{tab.ml.speedups}. 
For the system with $\epsilon_{\textrm{NP}} = 2, v=1$, the auto-tuning yielded a total simulated physical time of 4.5 ns compared to the 2.2 ns obtained with non-adaptive simulation. Figure \ref{ml.speedup.comparison} (inset) shows the variation in $\Delta$ as a function of the computational steps for the same systems. The tuning of $\Delta$ changed with the attributes of the ions and NP. Generally, longer $\Delta$ values (and associated higher speedup) were obtained for systems exhibiting a weaker dielectric contrast. 

Table \ref{tab.ml.speedups} quantifies the speedup by showing the performance comparison of the aforementioned ion-NP systems for 2 million steps on 4 MPI nodes each with 16 OpenMP threads and a fixed walltime of $\approx 20$ hours. ML-based adaptive timestep tuning enabled the simulation of the system for a longer physical time compared to the non-adaptive case under fixed compute resources and walltime. For the non-adaptive case, the virtual parameters $\mu = 30$ and $T_v = 0.002$ were selected using ML model I. A maximum speedup of $3.15$ was achieved for the ion-NP system defined with the input parameters: $\epsilon_{\textrm{NP}} = 60$, $\epsilon_\textrm{S} = 78.5$, $Q = -100$ , $v = 1$  and $M = 1272$, without adjusting any MPI or OpenMP parameters.

\begin{table}[htbp]
\caption{Performance comparison of different ion-NP systems simulated for 2 million steps ($\approx$ 20 hrs walltime) on 4 MPI nodes each with 16 OMP threads.}
\begin{center}
\begin{tabular}{|c|c|c|c|c|}
\hline
\multicolumn{1}{|c|}{\textbf{Physical System}}&\multicolumn{1}{|c|}{\textbf{Time (ns)}}&\multicolumn{1}{|c|}{\textbf{Speedup}}&\multicolumn{1}{|c|}{\textbf{Stability}} \\
\hline
\multicolumn{4}{|c|}{\textbf{Non-adaptive}}\\
\hline
2, 78.5, -100, 1 & 2.18& 1.00x& 0.01552  \\
\hline
\multicolumn{4}{|c|}{\textbf{ML-based Adaptive}}\\
\hline
2, 78.5, -100, 1 & 4.56& 2.09x& 0.00048 \\
\hline
2, 78.5, -100, 2 & 2.75& 1.26x& 0.00088 \\
\hline
40, 78.5, -100, 1 & 6.12& 2.81x& 0.00076 \\
\hline
60, 78.5, -100, 1 & 6.85& 3.14x& 0.00063 \\
\hline
\end{tabular}
\label{tab.ml.speedups}
\end{center}
\end{table}

\begin{figure}[tbh]
\centerline{\includegraphics[scale=0.84]{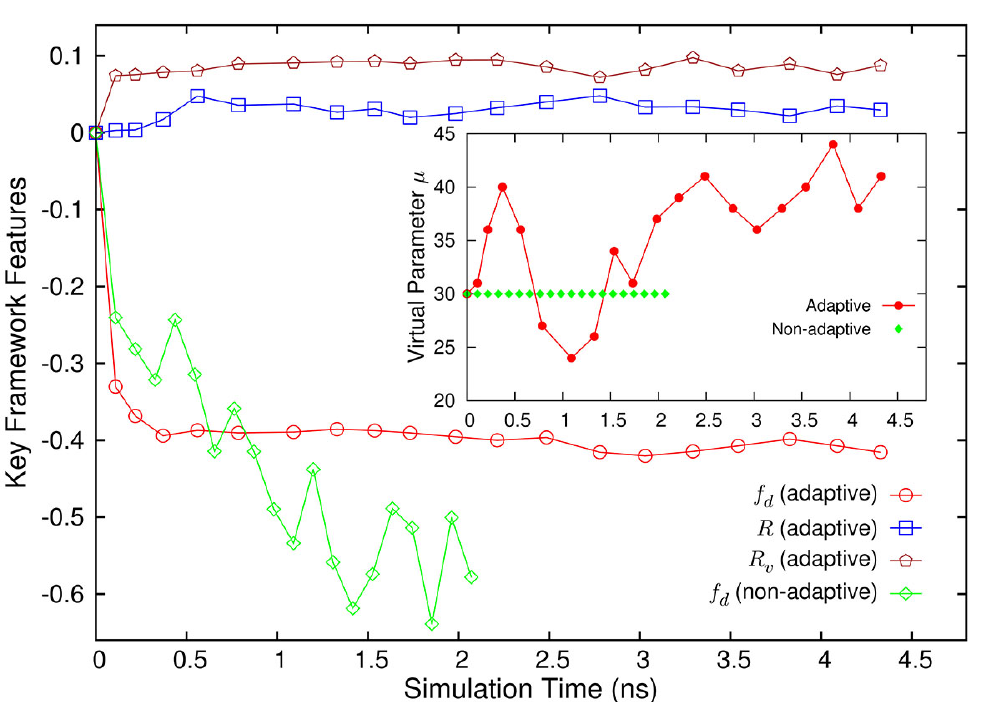}}
\caption
{\label{fig.stability.comparison}
Key output features $R, f_d, R_v$ as a function of the simulated physical time $t$ for the counterion-NP system characterized with NP charge $Q = -100$, ion valency $v = 1$, NP permittivity $\epsilon_{\textrm{NP}} = 2$, solvent permittivity $\epsilon_{\textrm{S}}= 78.5$, and mesh size $M = 1272$. ML-enabled auto-tuning of timestep $\Delta$ and virtual parameters produces enhanced stability (diminished fluctuations in $f_d$) compared to the non-adaptive case (stronger fluctuations in $f_d$). (Inset) ML-enabled auto-tuning of the virtual parameter $\mu$ for the same system (closed circles) and fixed $\mu$ for the non-adaptive case (closed diamonds).
}
\end{figure}

The ML-enhanced framework with adaptive timestep also improved the overall stability of the MD simulation. Figure \ref{fig.stability.comparison} shows the key output features associated with the simulation of the ion-NP system characterized with $\epsilon_{\textrm{NP}} = 2, v=1$ (other input parameters being the same as in Fig. \ref{ml.speedup.comparison}). Fluctuations in the output feature $f_d$, which measures the deviation of the on-the-fly optimized functional from the statically optimized functional, illustrate the stability of the simulation. Auto-tuning of parameters produced diminished fluctuations in $f_d$ compared to the non-adaptive framework. Figure \ref{fig.stability.comparison} (inset) shows the variation of virtual parameter ($\mu$) with simulated physical time for the same ion-NP system. By definition, the non-adaptive model produced constant $\mu$. On the other hand, the adaptive model produced the auto-tuning of $\mu$ which was correlated with the more stable dynamics (red circles characterizing $f_d$ in the outset of Fig. \ref{fig.stability.comparison}). Indeed, the variance in $f_d$ data for the non-adaptive model ($ f_{d | \sigma^2} = 0.01552$) was much higher than that for the adaptive model ($ f_{d | \sigma^2} = 0.00048$). Table \ref{tab.ml.speedups} shows the variance of $f_d$ for the same ion-NP systems analyzed above; in all cases, the variance was found to be significantly smaller by over an order of magnitude (indicating higher stability) for the adaptive model compared to the non-adaptive case. This enhanced stability can be attributed to the optimal updates of parameters $\mu, T_v$ during the intermediate times of the simulation. 

\begin{figure}[th]
\centerline{\includegraphics[scale=0.84]{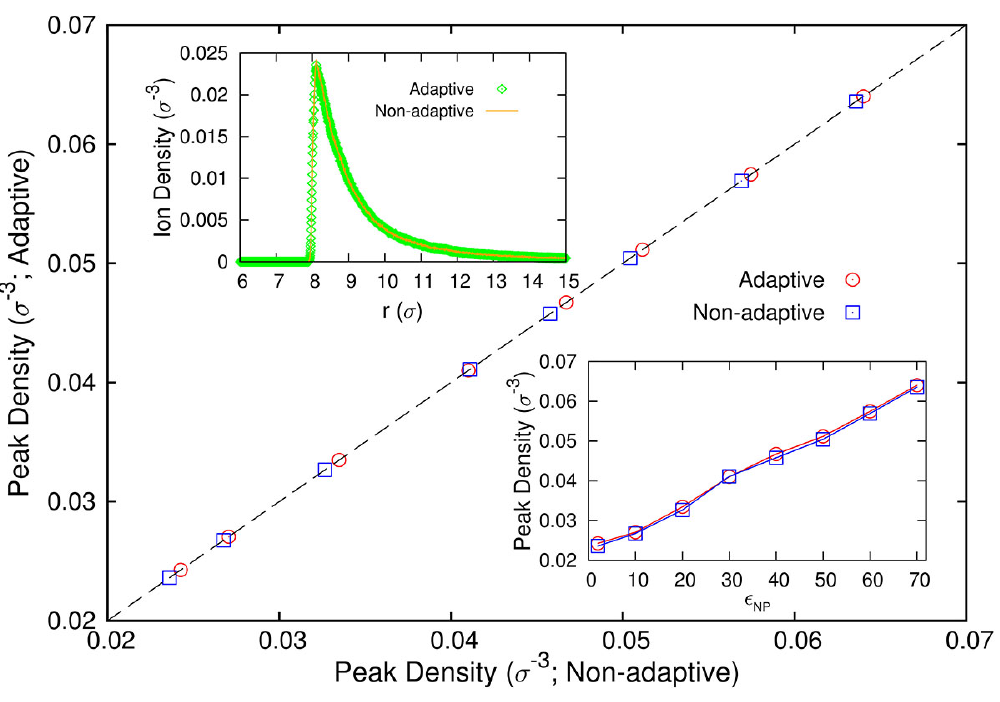}}
\caption
{\label{fig.accuracy}
Correlation between the peak densities associated with the distribution of counterions for ion-NP systems characterized by different NP permittivity $\epsilon_{\textrm{NP}}$; other parameters are the same as listed in the caption of Figure \ref{fig.stability.comparison}. Blue squares are values from the non-adaptive simulation, red circles are results from the ML-enabled adaptive simulation. (Top-left inset) Density distribution of counterions for the system with $\epsilon_{\textrm{NP}} = 2$; symbols are the results from the adaptive model, line corresponds to the non-adaptive case. 
(Bottom-right inset) Peak densities from the two models as a function of $\epsilon_{\textrm{NP}}$.
}
\end{figure}

In addition to increasing the efficiency and stability of the simulation, the framework with ML-enabled auto-tuning of parameters (adaptive framework) retained the accuracy associated with the framework using non-adaptive timestep and virtual parameters (non-adaptive framework). The accuracy can be assessed by comparing the density profiles of the ions computed using the two approaches. For different $\epsilon_{\textrm{NP}}$ values (other input parameters same as above), the peak densities computed using simulations based on adaptive framework were found to be in agreement with those calculated using the non-adaptive framework as shown in Figure \ref{fig.accuracy}; data from either approach falls on the dashed line which indicates linear correlation. Top-left inset of Figure \ref{fig.accuracy} shows the variation of the counterion density (in $\sigma^{-3}$, where $\sigma = 0.357$ nm is the ion diameter) as a function of the distance from the NP (of radius $7.5 \sigma$) for the specific case of $\epsilon_{\textrm{NP}} = 2$. The density profiles extracted from the adaptive and non-adaptive frameworks were found to be in good agreement (relative error in either distributions was found to be $\approx 1\%$). Bottom-right inset of Fig.~\ref{fig.accuracy} shows the variation of the peak density of counterions as a function of the dielectric permittivity $\epsilon_{\textrm{NP}}$ of the NP. Both approaches yield similar peak densities. Lowering $\epsilon_{\textrm{NP}}$ leads to an increase in the repulsive force on the counterions due to the induced charges on the NP surface, leading to the reduction in the peak density of counterions near the NP surface. 

\subsection{Benchmarking ML-enhanced Simulations}
The enhanced dynamical optimization framework was benchmarked using BigRed2 cluster nodes. 
These nodes have maximal achieved performance of 596.4 teraFLOPS, and feature a hybrid architecture based on two Cray, Inc., 344 XE6 (CPU-only) compute nodes, providing a total of 1,020 compute nodes, 21,824 processor cores, and 43,648 GB of RAM. Each XE6 node has two AMD Opteron 16-core Abu Dhabi x86$\_$64 CPUs and 64 GB of RAM; each XK7 node has one AMD Opteron 16-core Interlagos x86$\_$64 CPU, and 32 GB of RAM.

\begin{figure}[bth]
\centerline{\includegraphics[scale=0.84]{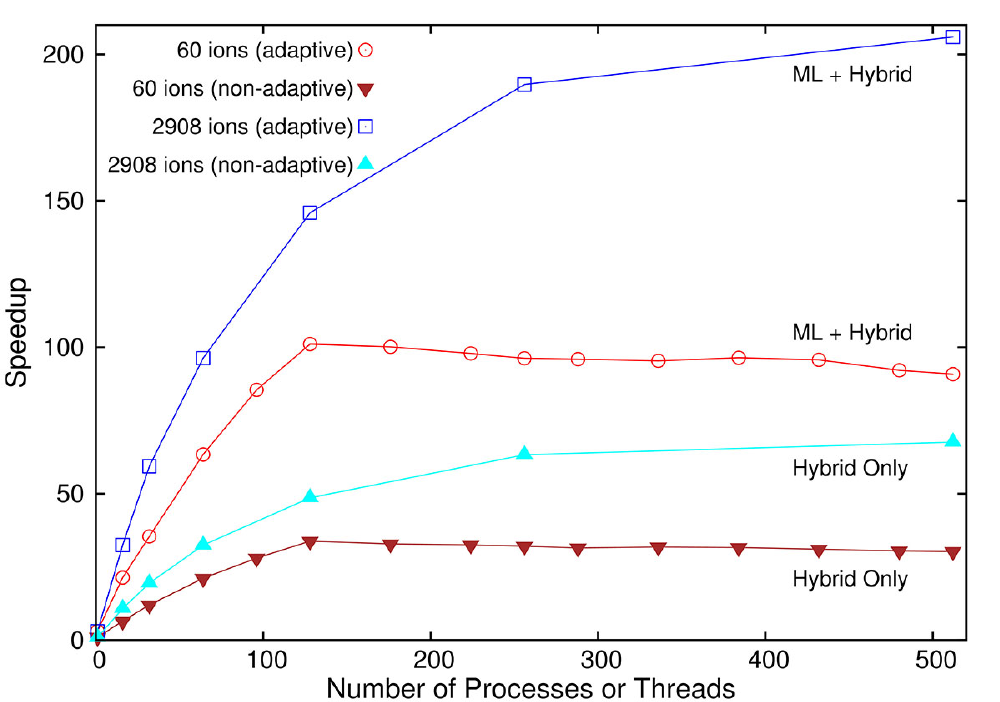}}
\caption
{\label{mpi.hybrid.scaling}
Strong scaling plot of the performance of the OpenMP/MPI hybrid technique with ML-enabled auto-tuning of the simulation timestep (open symbols) compared to the case with no auto-tuning (closed symbols). Data is shown for 60 and 2908 ions; in both systems, NP is meshed with 1082 mesh points (induced charges). For both systems, the combined ML and hybrid method outperforms the hybrid-only implementation.}
\end{figure}

Figure \ref{mpi.hybrid.scaling} compares the strong scaling plot of the performance of the adaptive and the non-adaptive dynamical optimization framework parallelized using the OpenMP/MPI hybrid model. The reported speedup is defined as the ratio of serial run-time to the time taken by the parallelized simulations (with and without ML-enabled auto-tuning) to simulate a set $t_b$ nanoseconds of ionic dynamics. In Figure \ref{mpi.hybrid.scaling}, results are shown for simulation of a system of 60 ions and 1082 mesh points as well as a larger system of 2908 ions and 1082 mesh points for $t_b \approx 2$ nanoseconds. The larger system of 2908 ions is comprised of both positive and negative ions characterized by an electrolyte concentration $c \approx 0.2$ M (with 60 counterions, 1424 positive electrolyte ions, and 1424 negative electrolyte ions). While the ML models were not trained with $c$ as an input parameter, simulations of the electrolyte systems using ML-predicted timestep and optimal virtual parameters associated with counterion-only systems were successful for up to $c\approx 0.4$ M (with varying $t_b$; see Section \ref{results.app}), enabling the benchmarking of simulations of larger systems. 

With non-adaptive timestep, the hybrid model produced the maximum speedup of 33.80 with 128 processes (8 MPI nodes and 16 OpenMP threads inside each MPI node) for the smaller system. For the same configuration, the hybrid model with ML-enabled auto-tuning of the simulation timestep was able to achieve a maximum speedup of 101.07 with 128 processes. Thus, the runtime for this system was reduced from 55 hours to 30 minutes (68 minutes without adaptive time-stepping). We note that the maximum speedup was calculated without considering the execution time reduction gained from the memory optimization techniques. For the hybrid model, we found that the optimal configuration of OpenMP threads is socket bound as noted in the literature \citep{rabenseifner2009hybrid}. As a result, the number of optimal OpenMP threads in our experiment was 16 for any number of MPI processes. 

When implemented to a larger system with a total number of 2908 ions and 1082 mesh points exhibiting induced surface charges, the combination of the hybrid methodology and the ML-based selection of adaptive timestep reduced the execution time of simulation for 2 nanoseconds from 88 days to 15.3 hours (32 hours without adaptive time-stepping) with a speedup of over 200. Clearly, the optimum number of MPI processes are proportional to the problem size when OpenMP thread affinity is set to the socket resulting in a well weak scaling system. The maximum speedup of 620.76 was obtained for 1024 processes executing a simulation of 5816 ions and 1272 mesh points for $t_b \approx 0.5$ nanoseconds.

\subsection{Application: Concentrated Electrolytes near an Oil-Water Emulsion Droplet}\label{results.app}
The ML-enhanced framework was applied to compute the distribution of monovalent electrolyte ions outside a charged oil-in-water emulsion droplet \citep{graaf,bier} at room temperature $T = 298$ K. Positive and negative ions were considered to be of the same size to simplify the system and focus on analyzing the effects of polarization charges on the density distributions. Such model systems have been considered in previous numerical studies of electrolyte ions near polarizable nanospheres \citep{messina1,santos,shen2017surface}. All ions were modeled as Lennard-Jones (LJ) spheres of diameter $\sigma = 0.357$ nm. The oil-water emulsion droplet was modeled as a spherical, dielectric interface with surface charge $Q = -60 e$ and radius $a=7.5\sigma \approx 2.7$ nm. The whole system of ions and the droplet was taken to be in a large spherical simulation cell of radius $b = 40\sigma \approx 14$ nm. The emulsion surface and the simulation cell boundary were modeled as spherical LJ walls. All excluded-volume interactions were modeled using the repulsive 6-12 LJ potential with $\epsilon_{\textrm{LJ}} = 1$ $\boltzmann T$ and cutoff $r_c = 2^{1/6}\sigma$. 

The dielectric permittivity of oil was taken to be $\epsilon_{\textrm{o}} = 2$, while water was associated with $\epsilon_{\textrm{w}} = 78.5$. The difference in the polarizable properties of oil and water lead to induced charges on the oil-water interface. Electrostatic interactions arising from the bare and induced charge interactions in the system were modeled using the forces originating from the functional described in Eq.~\eqref{eq:dsfnal}. The oil-water dielectric interface (which can be considered as the surface of the NP) was meshed with $M=1082$ mesh points; higher $M$ values were found to yield similar densities indicating that $M=1082$ was large enough to obtain converged density profiles. In addition to the system with no added electrolyte, systems with electrolytes characterized by concentration $c \approx 0.02, 0.1$ M were considered to analyze the effects of changing $c$ on the ionic distributions. Together with the 60 counterions (associated with the charged oil-water droplet), these concentrated electrolytes correspond to systems with a total of $350$ and $1514$ ions respectively. Simulation of the smallest system ($60$ ions) assuming non-polarizable NP surface was also performed for assessing the role of surface polarizability.

It should be noted that the ML procedure was only trained for smaller systems with counterions ($c = 0$, thus in the absence of co-ions). Further the training was performed for relatively smaller computational time (up to 2 million steps). With the application of the ML-enhanced framework to the aforementioned electrolyte systems, we are elucidating the transferability of the features learned for the smaller system to different, larger physical systems (with additional co-ions and salt counterions, and long-time dynamics). Such an extended application of the developed ML models is possible, in part, because the addition of electrolytes weaken the effective interaction between counterions and oil-water surface as a result of the screened electrostatic forces.

The aforementioned attributes of the physical system supply the input parameters for the enhanced dynamical optimization framework. Following the process elucidated in Fig. \ref{fig.overview}, these input parameters were first passed to the ML model I to predict the required virtual system parameters to kickstart the simulation. Two protocols were followed: in one case, the auto-tuning using ML model II was performed for the entire duration of the simulation by repeating the optimal timestep and virtual parameter pattern inferred by the ANN for 2 million steps interval for subsequent cycles of 2 million steps. In the other approach, ML model II was employed to auto-tune the timestep and virtual parameters up to the first 2 million steps and subsequent evolution was performed with fixed values of these parameters predicted by ML model I (non-adaptive). Both approaches yielded the same results for the densities within the error bars. The total number of steps $S$ were selected based on the convergence of the density distributions; $S$ was system-dependent and converged results were obtained after $\approx 7 - 20$ nanoseconds depending on the electrolyte concentration. 

\begin{figure}[t]
\centerline{\includegraphics[scale=0.84]{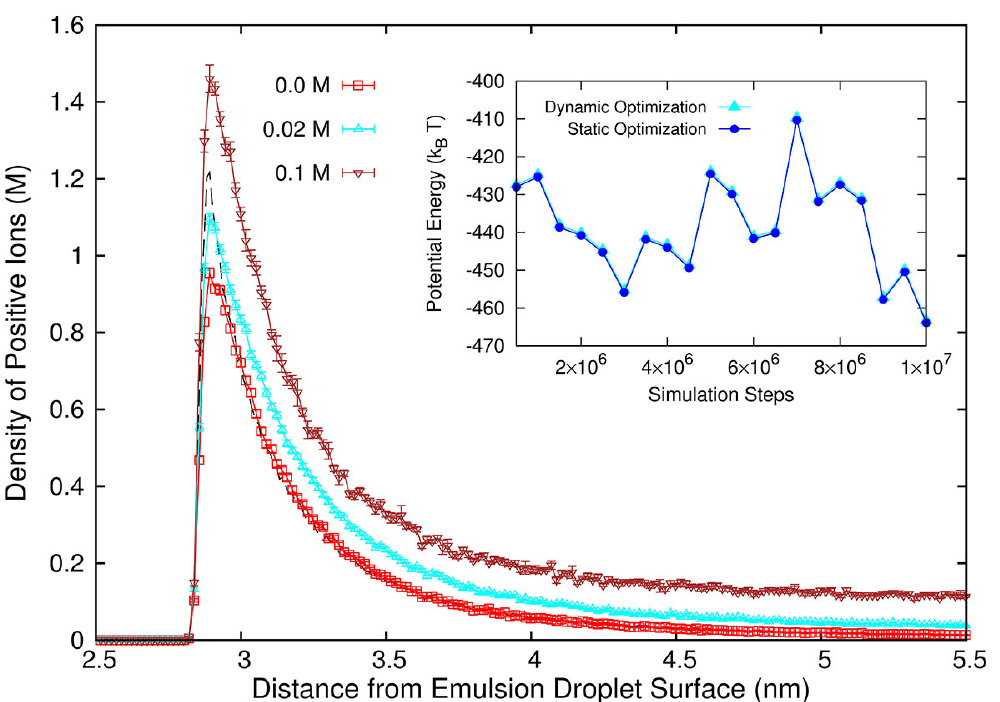}}
\caption
{\label{punchline.fig}
Ionic density profiles extracted from ML-enhanced MD simulations based on the dynamical optimization framework. Outset shows the density of positive ions for electrolytes of concentration $c \approx 0.0, 0.02, 0.1$ M near a negatively-charged ($-60 e$) oil-water emulsion with 60 associated counterions. The dielectric permittivity of oil and water is 2 and 78.5 respectively. Black dashed line refers to the result for the emulsion assumed to be unpolarizable (with oil permittivity as 78.5) at $c = 0$ M. (Inset) Comparison of the functional optimized dynamically (triangles) and the functional optimized at regular intervals keeping the ionic configuration static during the optimization process (circles) for the $c \approx 0.02$ M system.
}
\end{figure}

In all simulations, regardless of system sizes and presence/absence of electrolytes, good energy conservation was observed with $R < 0.05$. Similarly, the induced charges on the oil-water dielectric interface were accurately tracked by the on-the-fly optimization framework ($f_d < 1\%$; inset in Figure \ref{punchline.fig} shows the accurate tracking of the functional for the $c= 0.02$ M case). These two key features demonstrated the success of the ML-based virtual parameter selection process. As noted before, the bound on $R_v$ is system-size and system-feature-dependent; for counterion-only systems, $R_v < 0.15$ was recorded as per the expected limit while for electrolyte systems the bound on $R_v$ was higher and increased with increasing $c$. These findings demonstrate that ML models could be trained on smaller systems and applied to larger systems to obtain efficient and stable dynamics of ions in the latter case. 

Figure \ref{punchline.fig} shows the density profiles of the positive ions associated with the aforementioned systems. For all concentrations, the densities reach a constant value in the bulk away from the polarizable oil-water surface (negative ion densities, not shown, also reach a constant value in the bulk). Positive ions are found to accumulate near the dielectric interface, with the peak density increasing with $c$. Negative ions are depleted near the interface due to the repulsion from the bare charge on the oil-water surface as well as the induced charge. Comparison of the no electrolyte (counterion-only) result with the case where the surface is considered to be unpolarizable (with a permittivity equal to that of water) is also shown. The polarization charges on the surface lead to depletion of ions from the interface. Increasing electrolyte concentration leads to the rise in the peak density; the overall behavior is determined by the competition between the ion-induced charge repulsion near the surface and the ion-ion electrostatic and steric correlations. The position of the peak density remains relatively unaltered regardless of the $c$ value, as observed previously for monovalent electrolytes \citep{messina1}.

The inset in Fig. \ref{punchline.fig} shows the comparison of the potential energy functional optimized dynamically with the energy obtained after optimizing at regular intervals keeping the ionic configuration static during the optimization process for the $c \approx 0.02$ M system. The stability and accuracy evident from this plot demonstrates the success of the ML-based parameter selection process; the potential energy and the associated induced charges it characterizes were accurately tracked at all times for up to 10 million steps. Other systems showed similar agreement between dynamically optimized and statically optimized energy functionals. 

\section{Conclusion and Outlook}
We illustrated the computational gains accessible by integrating ML methods for parameter auto-tuning in MD simulations by demonstrating the enhancement of stability and efficiency of MD simulations of ions near polarizable NPs based on the dynamical optimization framework. The ANN-based ML model yielded the highest success rate among the non-linear regression models employed to predict the virtual system parameters at the start of the simulation. When integrated with the MD simulation, the ANN model predicted the timestep and the associated optimal virtual parameters with $94.3\%$  success rate. The auto-tuning of the simulation timestep resulting from the ML-enhanced, adaptive simulation framework enabled the simulation of ions for a longer physical time with a net speedup of $\approx 1.25 - 3$ (depending on the system configuration) compared to the non-adaptive simulation model. The combination of the ML procedure with the hybrid parallelization method generated stable dynamics of thousands of ions in the presence of polarizable NPs with computational time reducing from thousands of hours to tens of hours yielding a maximum speed up of $\approx 600$. Compared to the non-adaptive simulation with static initial virtual parameters, the stability of the adaptive framework increased by over an order of magnitude. 

This enhanced simulation framework has many applications and we demonstrated its utility by generating stable, accurate dynamics of ions in the presence of a polarized nanoemulsion droplet for up to $\approx 10$ million simulation steps. Additionally, we showed the broad applicability of the approach by demonstrating that the ML models trained on a smaller system can be applied successfully to produce accurate and stable dynamics of larger systems characterized by new attributes such as electrolyte concentration. At the same time, the approach reveals a limit on the electrolyte concentration and physical time that one can simulate based on training a counterion-only system. Future efforts will involve exploring the training of ML models with electrolyte concentration and attributes as input parameters. We will also explore integrating the current ML-enabled enhanced framework with fast Ewald solvers to support periodic boundary conditions and reduce the $O(n^2)$ scaling in system size to $O(n \log n)$ \citep{marchi,luijten.jcp2014}. 
Further, future work will involve exploring the training and design of ML models to predict the virtual parameters that generate stable ion dynamics near NPs of different shapes (e.g., discs, rods) \citep{brunk2019computational} as well as in systems where multiple dielectric interfaces are present \citep{shen2017surface}.

The use of ML to enhance the simulation framework enables users across the globe, with a diversity of domain experience, to simulate ions near polarizable NPs via the use of web-based applications hosted on services like nanoHUB. A tool powered by this enhanced framework was recently published on nanoHUB \citep{kadupitiya2018}. We will extend our framework to enable the process of using the data generated by this nanoHUB application for continuous training of the ML-based parameter prediction procedure. The approach presented here to integrate ML-based parameter prediction methods with MD-based simulations can be extended to other energy minimization problems \citep{brunk2019computational}. The implications of ML-enabled tuning of simulation timestep also suggest new avenues of exploring ML to advance simulation methods, such as ML-informed dynamic grid sizes in mesh-based problems.





\end{document}